\documentstyle[epsf,multicol,aps]{revtex}

\def\l{\langle}
\def\r{\rangle}

\begin{document}
\draft
\title{
Finite-size scaling for the Ising model on the M\"obius strip and 
the Klein bottle
}

\author{Kazuhisa Kaneda\cite{kaneda} and Yutaka Okabe\cite{okabe}}
\address{
Department of Physics, Tokyo Metropolitan University,
Hachioji, Tokyo 192-0397, Japan
}

\date{Received 22 June 2000}

\maketitle

\begin{abstract}
We study the finite-size scaling properties of the Ising model 
on the M\"obius strip and the Klein bottle. 
The results are compared with those of 
the Ising model under different boundary conditions, 
that is, the free, cylindrical, and toroidal 
boundary conditions.  
The difference in the magnetization distribution function 
$p(m)$ for various boundary conditions is discussed in terms of 
the number of the percolating clusters and the cluster size.
We also find interesting aspect-ratio dependence 
of the value of the Binder parameter at $T=T_c$ 
for various boundary conditions. 
We discuss the relation to the finite-size correction 
calculations for the dimer statistics. 
\end{abstract}

\begin{multicols}{2}
\narrowtext

The systems under various boundary conditions have the 
same per-site free energy in the bulk limit, whereas the 
finite-size corrections are different.  The idea of 
finite-size scaling (FSS) \cite{fisher70,cardy88}
is very important in understanding finite-size effects 
near the criticality.  
It is known that the FSS functions depend on the boundary 
conditions.  The difference in the FSS functions for the 
Ising model under the periodic and free boundary conditions 
has been discussed in connection with the universal FSS 
for the percolation problem \cite{hlc95} 
and the Ising model \cite{ok96}. 
The systems with the tilted boundary conditions have 
been also studied \cite{okkh99,zlk99}. 
Quite recently, Lu and Wu \cite{Lu99} have studied 
dimer statistics on the M\"obius strip and the Klein 
bottle.  These two systems are other examples of 
interesting boundary conditions, and their topological 
property is unique. 
The dimer statistics on the M\"obius strip has been also 
studied by Brankov and Priezzhev \cite{brankov93}.
The ground-state entropy of 
the Potts antiferromagnet on the M\"obius strip has been 
investigated \cite{shrock99}. 
In two dimensions (2D) the relevance of the finite-size properties 
to the conformal field theory is another source of 
interest \cite{cardy87,bcn86,affleck86}. 

\vspace{2mm}
\begin{figure}
\centerline{\epsfxsize=\linewidth \epsfbox{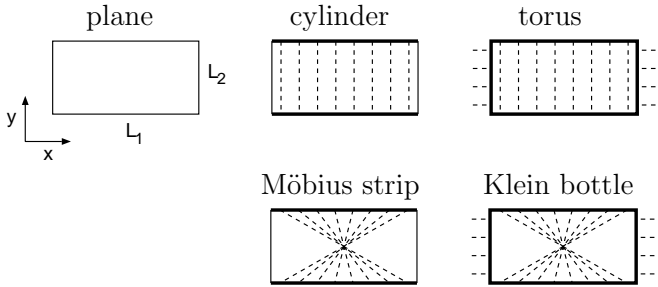}}
\vspace{4mm}
\caption{Illustration of the rectangular lattice with various 
 boundary conditions;  plane, cylinder, torus, M\"obius strip, and 
 Klein bottle.   Thick and thin lines denote 
 periodic and free boundaries, respectively.}
\label{boundary}
\end{figure}
\vspace{2mm}
In this Letter we study the FSS functions for the 2D Ising model 
on the M\"obius strip and the Klein bottle in view of 
increasing interest on the effect of boundary conditions 
for finite systems.  We compare these results with those of 
the Ising model under different boundary conditions, 
that is, the free, cylindrical, and toroidal 
boundary conditions.  As a total five boundary conditions 
are considered, and they are illustrated in Fig.~\ref{boundary}. 
We deal with the rectangular lattice of size $L_1 \times L_2$ 
with the aspect ratio $a=L_1/L_2$. 
Thick and thin lines denote periodic and free boundaries, 
respectively, in Fig.~\ref{boundary}. 
We impose the periodic boundary conditions in both the horizontal 
and vertical directions for the torus (toroidal boundary condition). 
The twisted periodic boundary condition is imposed 
in the $y$ direction, that is,
\begin{equation}
 f(x, y+L_2) = f(L_1-x, y), 
\label{twist}
\end{equation}
for the Klein bottle.  
The periodic boundary condition is imposed in one direction 
and the free one in the other direction for the cylinder 
(cylindrical boundary condition); 
the twisted periodic and the free boundary conditions 
are imposed for the M\"obius strip.  We refer to our system 
as a plane when we impose the free boundary conditions 
in both directions.  The M\"obius strip and the Klein bottle
have a unique topological property; they have 
a non-orientable surface.  The topological properties 
of five boundary conditions, the number of sides and the number of 
edges,  are tabulated in 
Table~\ref{topology}.  The symmetry property under 
a transformation $a \rightarrow 1/a$ is also given 
in Table~\ref{topology}.  The system is symmetric 
under such a transformation for the plane and the torus. 
\vspace{4mm}
\begin{table}
\caption{The topological properties of the rectangular lattices with 
various geometries (boundary conditions); the number of sides 
and the number of edge are given. 
The symmetry property under the transformation $a \rightarrow 1/a$ 
is also given.}
\label{topology}
\vspace{4mm}
\begin{tabular}{cccc}
geometry & \# of side & \# of edge & $a \rightarrow 1/a$ \\
\tableline
Klein bottle & 1  & 0 & no \\
M\"obius strip & 1  & 1 & no \\
torus & 2  & 0 & yes \\
cylinder  & 2  & 2 & no \\
plane  & 2  & 1 & yes \\
\end{tabular}
\end{table}

We use the Monte Carlo simulation to study the FSS properties 
of the 2D Ising model with various boundary conditions 
near the criticality.  
The moments of magnetization are basic quantities 
for the FSS analysis.  Here we focus on 
the Binder parameter \cite{binder81}, 
which is defined by
\begin{equation}
 g=\frac{1}{2} \Big(3-\frac{\l m^4 \r}{\l m^2 \r^2} \Big).  
\label{binder_ratio}
\end{equation}
One may determine the critical point from the 
crossing point of $g$ for different sizes as far as 
the corrections to FSS are negligible.  The value of the Binder 
parameter at the critical point is not a universal quantity, 
and it depends on the shape of the finite systems and the 
boundary conditions.  

\vspace{4mm}
\begin{figure}
\centerline{\epsfxsize=\linewidth \epsfbox{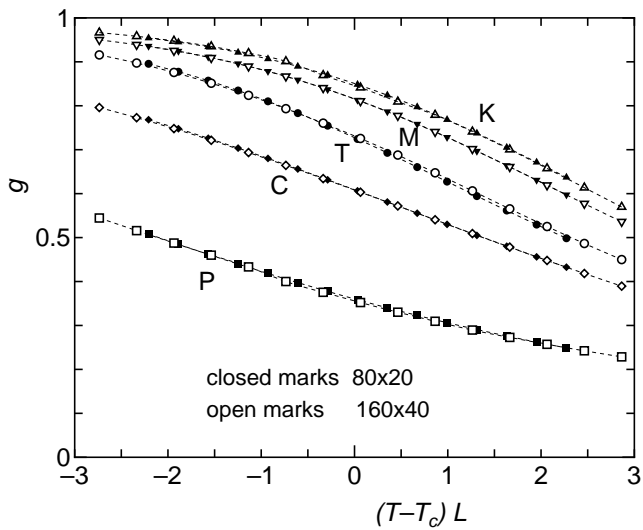}}
\vspace{4mm}
\caption{Plot of $g$ as a function of $(T-T_c) L^{1/\nu}$ 
 for the Ising model with various boundary conditions, 
 where $T_c=2.269 \cdots$ and $\nu$=1.  The data are plotted 
 in units of the coupling $J$.  
 Here, K, M, T, C, and P represent Klein bottle, 
 M\"obius strip, torus, cylinder, and plane, respectively. 
 The aspect ratio $a$ is chosen as 4.}
\label{g_klein}
\end{figure}
\vspace{2mm}
We plot the temperature dependence of the Binder parameter $g$ 
for several lattices with different boundary conditions 
in Fig.~\ref{g_klein}.  The system sizes are given within the 
figure, and the aspect ratio of the lattice is chosen as $a = 4$. 
The data of different sizes collapse on a single curve 
with using the scaling variable $(T-T_c) L^{1/\nu}$ 
for the horizontal axis.  
Here, $L=(L_1 \times L_2)^{1/2}$, $T_c = 2.269 \cdots$ and 
$\nu=1$ for the 2D Ising model; we plot the data in units of 
the coupling $J$. 
We have very good FSS behavior for $g$, and also find 
the strong dependence on the boundary conditions. 
The $g$ values of the Klein bottle are larger than those of 
the torus. The same behavior is found in the $g$ values of 
the M\"obius strip and those of the cylinder. 

\begin{figure}
\centerline{\epsfxsize=\linewidth \epsfbox{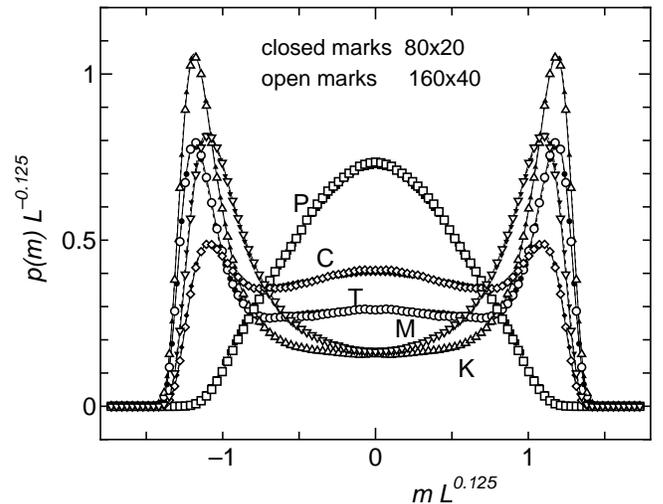}}
\vspace{4mm}
\caption{Plot of $p(m) L^{-\beta/\nu}$ at $T=T_c$ as a function 
 of $mL^{\beta/\nu}$ for the Ising model 
 with various boundary conditions, where $\beta/\nu$=1/8.  
 Here, K, M, T, C, and P represent Klein bottle, 
 M\"obius strip, torus, cylinder, and plane, respectively. 
 The aspect ratio $a$ is chosen as 4.}
\label{pm_klein}
\end{figure}
In order to clarify the difference in the Binder parameter $g$, 
we study the magnetization distribution function $p(m)$ at $T=T_c$. 
We show the FSS plots of $p(m)$ for various sizes 
with different boundary conditions in Fig.~\ref{pm_klein}. 
We plot $p(m) L^{-\beta/\nu}$ as a function of $m L^{\beta/\nu}$; 
$\beta = 1/8$ for the 2D Ising model. 
We again have good FSS behavior for $p(m)$ at $T=T_c$. 
The FSS function of $p(m)$ strongly depends on the boundary 
conditions.  There are two sharp peaks in $p(m)$ for the 
Klein bottle and M\"obius strip, and the secondary 
broad peak around $m=0$ becomes larger 
for the torus and cylinder. 
There is one peak around $m=0$ for the plane.  

\begin{figure}
\centerline{\epsfxsize=\linewidth \epsfbox{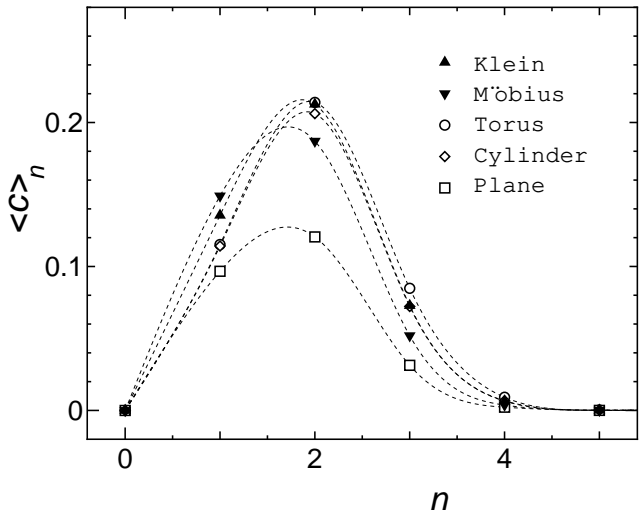}}
\vspace{4mm}
\caption{Plot of $\l c \r_n$ at $T=T_c$ as a function 
 of $n$ for the Ising model with various boundary conditions; 
 that is, Klein bottle, M\"obius strip, torus, cylinder, 
 and plane.  The system size is $160 \times 40$ ($a=4$).}  
\label{cn}
\end{figure}

We may understand this behavior from the number of 
percolating clusters and the size of clusters \cite{toh99}. 
For the systems with large aspect ratio, the importance 
of the number of percolating clusters has been pointed out 
by Hu and Lin \cite{hl96}. The probability for the appearance 
of $n$ percolating clusters for anisotropic lattices have been 
of current interest \cite{aizenman97,cardy98}. 
It has been revealed by Tomita {\it et al} \cite{toh99} that 
for the Ising model the combination of the percolating clusters 
with up spins and those with down spins gives the contribution 
to the broad peak around $m=0$ in $p(m)$. 
It is interesting to relate the effect of the boundary conditions 
to the number of percolating clusters and the size of clusters. 
The periodic boundary conditions may have the tendency that 
the size of the percolating clusters becomes larger 
compared to the free boundary conditions; the order will be reduced 
near a free surface.  Moreover, 
if one twists the boundaries, more clusters mix together.  
To confirm this speculation, we study the percolating properties 
of the Ising model with different boundary conditions. 
Using the fact that the Ising model is mapped to 
the percolation problem with the bond concentration of 
$1-e^{-2J/T}$ \cite{kf69,ck80}, 
we can assign clusters. 
Then, we may decompose the physical quantities 
by the number of percolating clusters \cite{toh99}.  
In Fig.~\ref{cn} we plot the fraction of lattice sites 
in the $n$ percolating clusters $\l c \r_n$ at $T=T_c$ 
for the system of size $160 \times 40$.  
We find from the figure that the cluster sizes become smaller 
for the system with free bondary conditions, and 
the single cluster is actually more dominant 
for the Klein bottle and the M\"obius strip. 
Thus, we explain the large $g$ values of the M\"obius strip and 
the Klein bottle in relation to the number of percolating clusters. 
\vspace{2mm}

\begin{figure}
\centerline{\epsfxsize=\linewidth \epsfbox{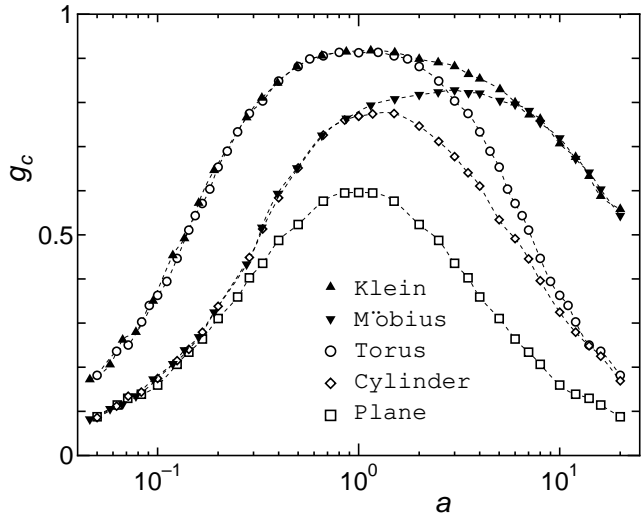}}
\vspace{4mm}
\caption{Aspect ratio dependence of the Binder parameter at $T=T_c$, 
 $g_c$, for the Ising model with various boundary conditions; 
 that is, Klein bottle, M\"obius strip, torus, cylinder, 
 and plane.}
\label{ga_klein}
\end{figure}

The FSS functions for the Binder parameter $g$ and 
the distribution function $p(m)$ depend on both the aspect ratio 
and the boundary condition.  We plot the aspect-ratio dependence 
of $g$ at $T=T_c$, $g_c$, for various boundary conditions in 
Fig.~\ref{ga_klein}.  We use the logarithmic scales for the 
horizontal axis.  We have several interesting observations 
from Fig.~\ref{ga_klein}.  The plot of $g_c$ as a function of $a$ 
in logarithmic scale is symmetric for the torus and the plane 
because of the symmetric property 
under the transformation $a \rightarrow 1/a$; $g_c$ takes the maximum 
at $a=1$. This is not the case for other geometries.  
For large enough $a$ ($\gg 1$), the FSS properties of the Klein bottle 
and those of the M\"obius strip become the same because 
the boundaries along the shorter direction determine the 
FSS properties of the system; for both the M\"obius strip and 
the Klein bottle, the boundary condition along $y$ axis is 
the twisted periodic one.  The FSS properties of the 
torus and the cylinder are the same for large enough $a$. 
In contrast, the systems with $a \le 1$, the Klein bottle and 
the torus show similar FSS behavior.  It is because the twisted 
periodic boundary or simple periodic one is not 
important for the number of percolating clusters for $a \le 1$. 
It is the same situation for the M\"obius strip and the cylinder 
for $a \le 1$.  For small enough $a$ ($\ll 1$), 
the M\"obius strip, the cylinder, and the plane show 
the same FSS properties because the boundaries along the 
shorter directions for these three are the same, that is, 
the free boundary condition.  

\vspace{2mm}
Let us compare our results with those 
of the dimer statistics by Lu and Wu \cite{Lu99}, 
who have calculated the finite-size corrections for 
the dimer generating function.  The finite-size 
correction coefficients have been tabulated in Table 1 
of Ref.~\cite{Lu99}.  Their notation $M \times N$ 
for the system size corresponds to our $L_1 \times L_2$. 
In the limit of large $M$ in their notation, 
which corresponds to $a \gg 1$, the finite-size correction 
coefficients ($c_2, \Delta_2$) 
are classified into three groups, that is, M\"obius 
strip and Klein bottle, cylinder and torus, and plane. 
In contrast, for large $N$ in their notation, $a \ll 1$, 
the finite-size correction coefficients 
($c_1, \Delta_1$) are classified into two groups;  
one group is Klein bottle and torus, and the other is 
M\"obius strip, cylinder, and plane.  For both cases 
their results of the dimer statistics are consistent 
with the present results.  In addition, 
for large $M=N$, which corresponds to $a = 1$, 
the correction coefficients $(c_1, c_2)$ are relevant; 
then Klein bottle and torus, M\"obius strip and cylinder, 
and plane form three groups.  This result is again consistent 
with the present one. 
These observations are compatible with the conformal 
field theory \cite{cardy87,bcn86,affleck86} that 
the behavior of the difference in finite-size 
corrections for different boundary conditions 
are model-independent.

To summarize, we have studied the FSS properties of the Ising model 
with various boundary conditions, that is, 
Klein bottle, M\"obius strip, torus, cylinder, and plane. 
We have elucidated the difference in the magnetization 
distribution function $p(m)$ for various boundary 
conditions in terms of the number of the percolating clusters 
and the cluster size.
We have found interesting aspect-ratio dependence 
of the $g$ value at $T=T_c$ for various boundary 
conditions.  
The FSS properties of the systems are classified into 
three groups for $a \gg 1$ and two groups for $a \ll 1$. 
For large $a$, the $g$ value becomes large 
for the M\"obius strip and the Klein bottle, which is 
characteristic for systems with a non-orientable surface.

There may be several directions for future study. 
In the dimer statistics, two identities relating dimer 
generating functions for M\"obius strips and cylinders 
have been established \cite{Lu99}.  It is desirable 
to explore exact relations between the Binder parameters 
of the Ising systems with different boundary conditions. 
Anyons on the cylinder and the torus have been studied with 
the braid-group analysis \cite{hatsugai91};  the topology 
of the systems is important in the fractional quantum 
Hall effect.  Anyons on the M\"obius strip and the Klein bottle 
will be interesting subjects to study.  
The boundary condition dependence of the critical behavior 
of the Anderson transition has been investigated \cite{slevin00}. 
It is also interesting to study the scaling functions for the 
Anderson transition with the M\"obius and Klein-bottle 
boundary conditions.

Several systems on the M\"obius strip and the Klein bottle 
have been studied in various fields of physics. 
Persistent currents in a M\"obius ladder have been 
studied \cite{mila98}, and interesting finite-size effects 
have been discussed.  A general construction of correlation 
functions in rational conformal field theory on the M\"obius strip 
and the Klein bottle has been made in terms of three-dimensional 
topological quantum field theory \cite{felder99}. 
Moreover, the matrix string theory has been constructed on the M\"obius 
strip and the Klein bottle \cite{zwart98}. 
The present work on the interplay of the topology and 
aspect ratio in the FSS properties of the Ising model 
may accelerate the study of physics on a non-orientable surface. 

After we submitted our manuscript, we came to know 
the preprint by Lu and Wu \cite{Lu00}.  They have studied 
the partition function of the Ising model on the M\"obius strip 
and the Klein bottle analytically.  They obtained essentially 
the same conclusion as that of the dimer statistics \cite{Lu99}. 
We have made the detailed study of the boundary-condition 
dependence of the magnetization distribution function $p(m)$, 
which is complementary to the analytical work on the partition 
function.

Thanks are due to N. Kawashima, Y. Tomita, and M. Kikuchi 
for valuable discussions. 
The computation in this work has been done using the facilities of
the Supercomputer Center, Institute for Solid State Physics,
University of Tokyo.
This work was supported by a Grant-in-Aid for Scientific Research 
from the Ministry of Education, Science, Sports and Culture, Japan.

\end{multicols}
\end{document}